\documentclass[journal]{new-aiaa}
\usepackage[utf8]{inputenc}
\usepackage{textcomp}

\usepackage{graphicx}
\usepackage{amsmath}
\usepackage[version=4]{mhchem}
\usepackage{siunitx}
\usepackage{longtable,tabularx}
\usepackage{multirow}
\usepackage{bigfoot}
\usepackage[dvipsnames]{xcolor}

\title{Semi-Analytical Model for Design and Analysis of On-Orbit Servicing Architecture}

\author{Koki Ho\footnote{Assistant Professor, Daniel Guggenheim School of Aerospace Engineering, Atlanta, GA, AIAA Senior Member}}
\affil{Georgia Institute of Technology, Atlanta, GA, U.S.A.}
\author{Hai Wang\footnote{Assistant Professor, School of Information Systems, Singapore, Singapore; Visiting Assistant Professor, Heinz College of Information Systems and Public Policy, Pittsburgh, PA, U.S.A.}}
\affil{Singapore Management University, Singapore, Singapore\\
Carnegie Mellon University, Pittsburgh, PA, U.S.A.}
\author{Paul A. DeTrempe\footnote{M.S. Student, Department of Aeronautics and Astronautics, Stanford, CA}}
\affil{Stanford University, Stanford, CA, U.S.A.}
\author{Tristan Sarton du Jonchay\footnote{Ph.D. Student, Daniel Guggenheim School of Aerospace Engineering, Atlanta, GA, AIAA Student Member}, and Kento Tomita\footnote{Ph.D. Student, Daniel Guggenheim School of Aerospace Engineering, Atlanta, GA, AIAA Student Member}}
\affil{Georgia Institute of Technology, Atlanta, GA, U.S.A.}

\begin{document}

\maketitle

\begin{abstract}
Robotic on-orbit servicing (OOS) is expected to be a key technology and concept for future sustainable space exploration. This paper develops a novel semi-analytical model for OOS system analysis, responding to the growing needs and ongoing trend of robotic OOS. An OOS infrastructure system is considered whose goal is to provide responsive services to the random failures of a set of customer modular satellites distributed in space (e.g., at the geosynchronous orbit). The considered OOS architecture comprises a servicer that travels and provides module-replacement services to the customer satellites, an on-orbit depot to store the spares, and a series of launch vehicles to replenish the depot. The OOS system performance is analyzed by evaluating the mean waiting time before service completion for a given failure and its relationship with the depot capacity. By uniquely leveraging queueing theory and inventory management methods, the developed semi-analytical model is capable of analyzing the OOS system performance without relying on computationally costly simulations. The effectiveness of the proposed model is demonstrated using a case study compared with simulation results. This paper is expected to provide a critical step to push the research frontier of analytical/semi-analytical model development for complex space systems design.
\end{abstract}

\section*{Nomenclature}

{\renewcommand\arraystretch{1.0}
\noindent\begin{longtable*}{@{}l @{\quad=\quad} l@{}}
$C$ & on-orbit depot spare capacity, in units of modules\\
$D$ & spare demand, in units of modules per hour\\
$L$ & lead time, in units of hours\\
$N$  & number of customer satellite modules, in units of modules\\
$S$ & service time, in units of hours\\
$S_{\text{inbound}}$ & inbound travel time, in units of hours\\
$S_{\text{outbound}}$ & outbound travel time, in units of hours\\
$S_{\text{repair}}$ & repair time, in units of hours\\
$S_{\text{stockout}}$ & extra service time due to stockout, in units of hours\\
$T_\text{l}$ & launch interval, in units of hours\\
$T_\text{s}$ & time until stockout, in units of hours\\
$W$ & waiting time until service completion, in units of hours\\
$W_\text{q}$ & waiting time in queue, in units of hours\\
$\alpha$ & mean individual module failure rate, in units of failures per hour\\
$\beta$ & mean launch rate, in units of launches per hour\\
$\lambda$ & mean system-level spare demand rate, in units of modules per hour\\
$\Phi$ & fill rate\\
$\Phi_{\text{req}}$ & fill rate requirement\\
\end{longtable*}}

{\renewcommand\arraystretch{1.0}
\noindent\begin{longtable*}{@{}l @{\quad=\quad} l@{}}
$f_{\cdot}$ & probability density function\\ 
$\mathbb{E}\left[ \cdot\right] $ & expected value\\
$\left\lbrace \mathcal{L}*\cdot\right\rbrace (\theta)$ & Laplace-Stieltjes transform of a function
\end{longtable*}}

\section{Introduction}

Nowadays, there has been increasing interest in developing on-orbit infrastructure systems that enable sustainable space exploration. Over the last several decades, research and development in autonomous and robotics systems have significantly raised the technology readiness level of robotic on-orbit servicing (OOS) \cite{NASA,JAXA,DARPA}. Engineers envision space-based servicing infrastructures to provide refueling and repair services or to manufacture large structures directly in orbit. The recent trend of satellite modularization is also enabling the concepts of “servicing-friendly” spacecraft that are composed of multiple small structural modules with standardized interface mechanisms \cite{Sullivan2001,Hill2013,Kerzhner2013,Jaeger2013}.
These OOS infrastructure systems and serviceable satellite designs are expected to be game-changing technologies for the satellite industry \cite{Saleh2002,Lamassoure2002,Saleh2003,Long2007,Saleh2001}.  

Traditionally, most OOS concepts studied in the literature have assumed a dedicated robotic spacecraft to repair or refuel the customer satellites \cite{Yao2013,Zhao2016,Verstraete2018}. The servicer would visit a predefined set of satellites and be discarded once the mission is over. The advantage of this concept is that the flight path of the servicer can be optimized before launch to provide the best value to the service operation. However, although such a “disposable" servicer may be a favorable concept in the near term, it is not a sustainable solution for OOS in the longer term. 

Alternatively, a more sustainable concept is to use a permanent reusable servicing infrastructure that responds to random failures \cite{Long2007,SartonduJonchay2017}. One example of such a concept would include a servicer replacing defective modules with new module spares, an on-orbit depot storing the spare modules, and a series of launch vehicles supplying new modules from Earth to the depot on a regular basis. When a module fails on a satellite, the OOS system dispatches a servicer with a new spare to that satellite to replace the failed module with the spare. This concept can provide timely and responsive services to the random failures spontaneously and thus enable sustainable space exploration. 

However, despite the potential advantage of a permanent responsive OOS infrastructure system, its design and operations planning are substantially more complex and challenging compared with the traditional dedicated servicer concepts. This is because the analysis of a responsive OOS system would need to consider the interactions between the infrastructure elements as well as the full supply chain of the spare modules, including the queue of the services and the inventory of the spare depot. We need an efficient model to analyze the design of such a complex OOS system.

In response to this background, we develop a novel semi-analytical model to evaluate the OOS system design and analysis. Given a set of distributed modular satellites (i.e., customer satellites) with random failure rates, the proposed model can analyze the performance of the servicing system considering the spares' supply chain. Specifically, it can evaluate the mean waiting time before service completion for a given customer satellite failure and its tradeoff relationship with the depot capacity.

The developed semi-analytical model uniquely leverages queueing theory and inventory management methods. Queueing theory is the mathematical theory of waiting in lines; it models real-world queueing systems using distributions of customer arrival, service time, and queue discipline \cite{LarsonOdoni}. Inventory management refers to the process of ordering, storing, and using inventory, such as raw materials, components, and end products; inventory control methods can be applied to find an order policy that balances inventory cost and demand shortage \cite{SimchiLevi}. Based on these theories, the developed model for the OOS system contains a set of coupled submodels: (1) a queueing submodel that models the mean waiting time before service completion for a module failure; and (2) an inventory control submodel that models the replenishment of the on-orbit depot from the ground. The results from the semi-analytical model are compared with simulation results for different real-world cases, and the accuracy of the proposed model is demonstrated. 

A few remarks need to be made about the value and contribution of the semi-analytical model developed in this paper. Conventionally, the analysis of space systems with random failures and repairs (e.g., OOS systems) has been performed using computationally costly discrete-event or agent-based simulations \cite{Baldesarra2007,Yao2013,Long2007,Richards,SartonduJonchay2017,Sears2018}. However, as space systems become complex, designing and evaluating their performance using only simulations becomes computationally challenging. While simulations are effective for detailed design, there is a growing need for the development of more efficient, yet rigorous, analysis methods to enable quick performance evaluation for systems design and trade space exploration. Although analytical or semi-analytical models have been recently introduced gradually in space systems design \cite{Jakob,Zchen}, this research direction remains largely unexplored. This paper introduces the first semi-analytical approach with an integrated queueing and inventory model for complex space systems analysis. 
The developed approach enables the evaluation of the OOS system performance without relying on computationally costly simulations, reducing the computational time from hours down to seconds. 
The model developed in this paper is expected to be a critical step in pushing forward this research frontier of analytical/semi-analytical model development for space systems design.

The rest of the paper is organized as follows. Section \ref{overview} discusses the overview of the considered OOS architecture. Section \ref{semi-analytical} explains the main contribution of the work: a semi-analytical model for the OOS system. Section \ref{simulations} assesses the results from the proposed model using simulations with a realistic application example, and Section \ref{conclusion} concludes the paper.

\section{Overview of On-Orbit Servicing Infrastructure}
\label{overview}
\begin{figure}[hbt!]
\centering
\includegraphics[scale=0.7]{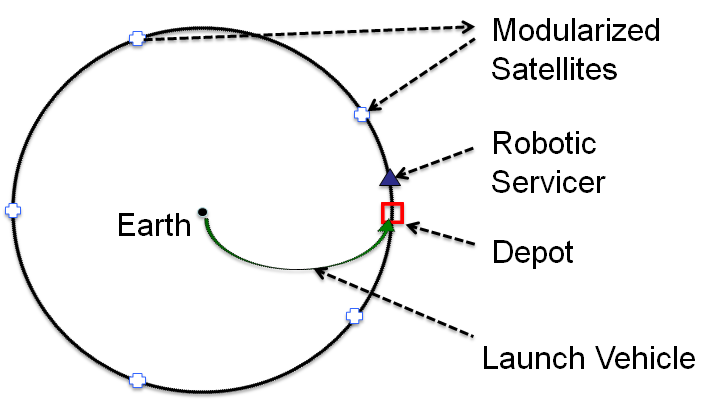}
\caption{Illustration of customer, servicer, and resupply architecture.}
\label{ServiceArchitecture}
\end{figure}

This paper considers an OOS architecture to provide module-replacement services to a set of customer satellites distributed in space (e.g., at the geosynchronous orbit). The customer satellites are modular, where each module can fail and be replaced independently. The servicing architecture comprises three main components: the servicer, the on-orbit depot, and the launch vehicles. The on-orbit depot stores modules which are brought to customers by the servicer when failures (i.e., demand for services) happen. The launch vehicles are used to refill the depot. An illustration of the considered servicing architecture is shown in Fig. \ref{ServiceArchitecture}.

The overview of the concept of operations is as follows. The servicer remains docked to the depot while awaiting a customer satellite module failure. Once a customer satellite module fails, the servicer performs a maneuver to the failed module. After the rendezvous and docking with the customer satellite, the servicer performs the repair operation, defined as the replacement of a failed module with a spare module, and then returns to the depot via another maneuver.
Customers receive services in the order in which they failed. If a failure happens while the servicer is busy, it has to wait until the servicer becomes available (i.e., completes all the previous services in the queue) to receive the service. The servicer can only carry one spare module, and thus it must return to the depot to load a new spare and, if needed, refuel itself prior to the next repair trip. The spare inventory in the depot is monitored regularly, and the launch vehicle visits the depot for its resupply following a stochastic launch schedule. 

\section{Semi-Analytical Model}
\label{semi-analytical}
This section provides an overview of the problem statement, the details of the developed model, and the proposed solution method based on the model.
\subsection{Problem Statement}
We develop a novel semi-analytical model to evaluate the OOS system design and analysis without relying on computationally costly simulations. As the first step, this subsection converts the concept of operations of the considered OOS system into a mathematical problem.

\begin{figure}[hbt!]
	\centering
	\includegraphics[trim=0 0 10 0,clip,scale=0.92]{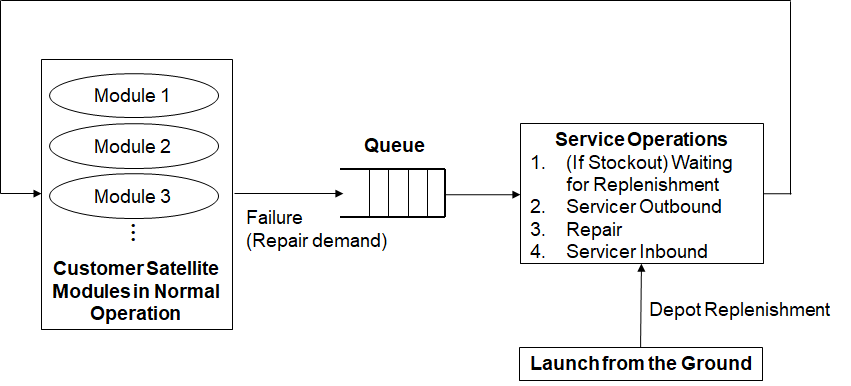}
	\caption{Schematic representation of the customer satellite modules and servicing architecture.}
	\label{BlockDiagram}
\end{figure}

The overall concept of operations can be expressed in a schematic diagram in Fig. \ref{BlockDiagram}. Following the architecture discussed in Section \ref{overview}, the model contains a set of customer satellite modules operated in orbit; when a failure happens to a module, it is added to the queue for service operations and is processed on a first-come-first-served (FCFS) basis. To sustain the service operations, the depot is replenished from the ground regularly. In this study, we consider all modules to be identical for simplicity, and the failures of each module are assumed to follow a (mutually independent) Poisson process. 

The queue for the services needs to be analyzed rigorously to capture both the service time and the waiting time. The service time is the time during which the servicer is dedicated to a service operation (i.e., from the moment when the servicer becomes available for a service operation until it returns to the depot upon completion of that service). The waiting time, on the other hand, is defined as the time that a customer satellite module failure (i.e., demand) has to wait until it is serviced (i.e., from the moment when a failure happens until its repair service is completed by the servicer). Mathematically, we can define the service time $S$ and the waiting time $W$ as follows:
\begin{equation}
\label{serviceTime}
S=S_{\text{stockout}}+S_{\text{outbound}}+S_{\text{repair}}+S_{\text{inbound}}
\end{equation}
\begin{equation}
W=W_\text{q}+S_{\text{stockout}}+S_{\text{outbound}}+S_{\text{repair}}%=W_\text{q}+S-S_{\text{inbound}}
\end{equation}
where $S_{\text{stockout}}$ is the delay of the service when the depot is found to be out of stock\footnote{Technically speaking, $S_{\text{stockout}}$ is not a service time, but we consider it as part of the service time so that the queueing and inventory control methods can be coupled naturally.}; $S_{\text{outbound}}$ and $S_{\text{inbound}}$ are the outbound and inbound (i.e., return) travel times; $S_{\text{repair}}$ is the repair time; and $W_\text{q}$ is the waiting time in the queue.
Here, $S_{\text{outbound}}$ and $S_{\text{inbound}}$ can be found from the (known) distribution of the positions of the customer satellites; these times can also include the time for the operations needed before and/or after each repair trip (e.g., loading a new spare, refueling). $S_{\text{repair}}$ is assumed to be a fixed value in this paper but can be varied if needed. The evaluation of the remaining terms requires an integrated queueing and inventory control analysis; $S_{\text{stockout}}$ is an output from the inventory control analysis, and $W_\text{q}$ is an output from the queueing analysis. Note that $W_\text{q}$ depends on the service times of all the previous repairs in the queue, and the length of the queue itself is also probabilistic. This coupled relationship makes the problem challenging.

\begin{figure}[hbt!]
    \centering
	\includegraphics[scale=0.7]{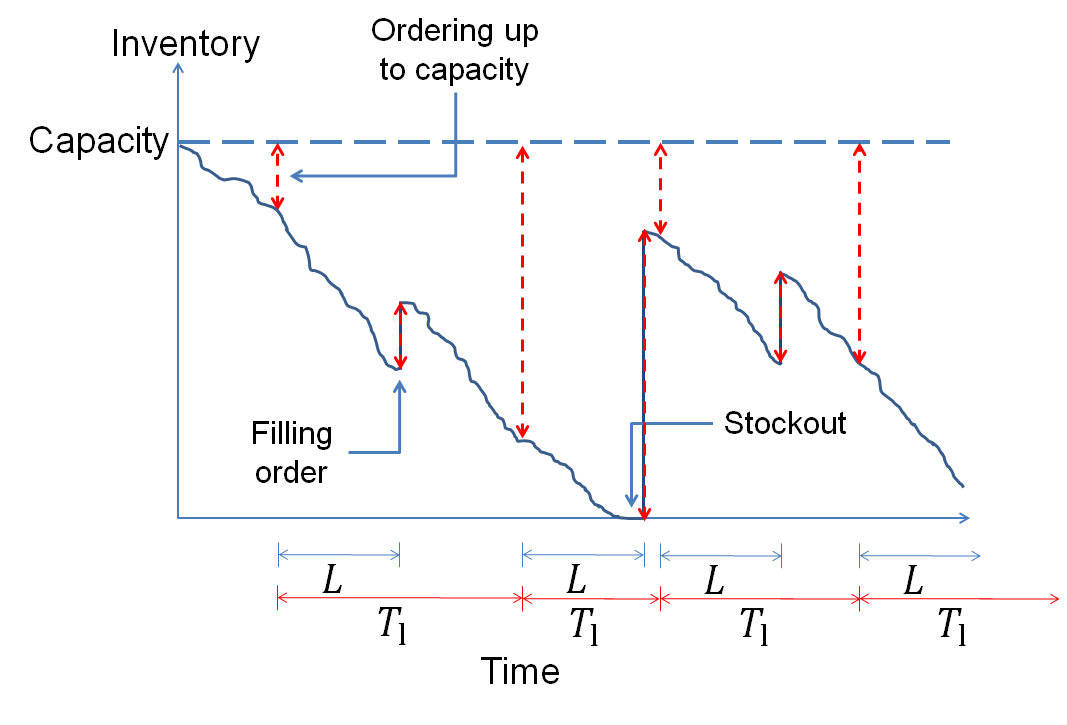}
	\caption{Order-up-to policy for the depot inventory control.}
	\label{inventory}
\end{figure}

The considered inventory control strategy of the depot can be modeled as a modified version of the order-up-to policy \cite{Cachon}. Namely, over every exponentially distributed time interval $T_\text{l}\sim\epsilon(\beta)$, we review the inventory and order $C-(I+I')$ units from the ground, where $C$ is the capacity of the depot, $I$ is the current inventory level, and $I'$ is the replenishment on the way (i.e., orders being processed). Here, the inventory level $I$ is defined as the number of units physically in stock minus the number of backorders, where the backorders are used to track the unmet demand due to stockout so that it can be delivered at subsequent opportunities. The review frequency is driven by the launch frequency of the rocket with a mean launch rate $\beta$. The exponential distribution for the launch interval has been shown to be a good approximation; see Appendix of Ref. \cite{Jakob}. A constant lead time $L$ is added between the review opportunity and the actual delivery of the spares in order to account for the processing of the order, the manufacturing of the units, the loading of the units onto the rocket, and the flight time to the depot.
Fig. \ref{inventory} summarizes the considered inventory control policy.

\begin{figure}[hbt!]
	\centering
	\includegraphics[scale=0.75]{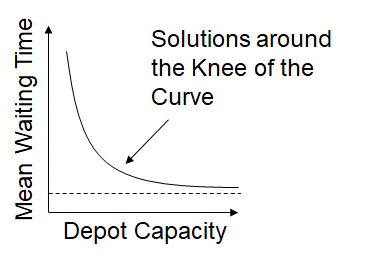}
	\caption{Solutions around the knee region of the curve of mean waiting time vs. depot capacity}
	\label{Knee}
\end{figure}

A key consideration for the inventory performance is the fill rate $\Phi$, representing the percentage of the spare demand that is met without stockout. We assume that we are given a fill rate requirement $\Phi_{\text{req}}$ (typically close to one), and our goal is to analyze the mean waiting time $\mathbb{E}\left[W\right]$ and the depot capacity $C$. The tradeoff between $\mathbb{E}\left[W\right]$ and $C$ corresponds to a cost-performance tradeoff. We prefer a low-cost and high-performance OOS system, which can be interpreted as a system with a small $C$ and a small $\mathbb{E}\left[W\right]$; however, as shown later, a smaller $C$ typically indicates a larger $\mathbb{E}\left[W\right]$, and thus the tradeoff between these two needs to be considered.
Particularly, we are interested in the behavior of $\mathbb{E}\left[W\right]$ vs. $C$ around the knee region of the curve\footnote{There are multiple definitions of the knee point/region in the multi-objective optimization literature \cite{Das,Sudeng}; in this paper, we use the concept of "the knee of the curve" to represent the region around the point of diminishing returns.}, which represents the depot capacity $C$ above which little additional saving is expected in the mean waiting time $\mathbb{E}\left[W\right]$. In practice, this indicates the solution point beyond which it is not worth investing more on expanding the depot capacity considering the small performance gain (i.e., waiting time reduction); the definition of that exact solution would depend on the cost model for the depot and the (monetary) penalty model for the waiting time, both of which are application-dependent. In this paper, instead, we aim to develop a general method to efficiently evaluate the solutions around that knee region rather than specifying the exact point. This concept is shown in Fig. \ref{Knee}. Note that this knee region corresponds to where $\Phi_{\text{req}}$ is close to one; therefore, in later numerical examples, we particularly focus on the solutions around $\Phi_{\text{req}}= 0.95$ or above, which is true for a realistic reliable OOS system.

\subsection{Proposed Model}
\begin{figure}[hbt!]
	\centering
	\includegraphics[scale=0.8]{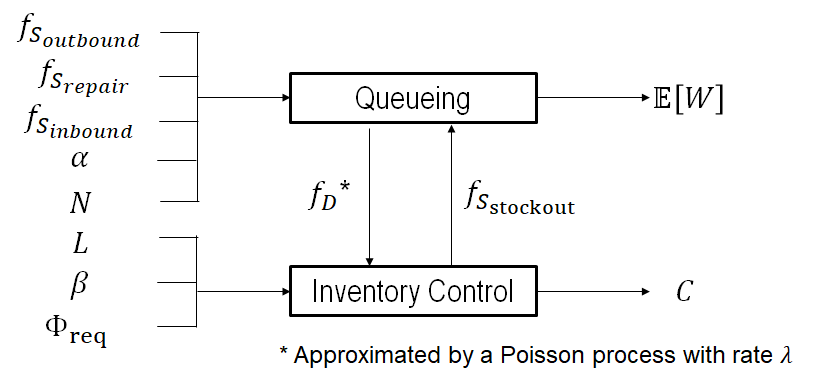}
	\caption{Integrated queueing and inventory model for the OOS system.}
	\label{Integrated}
\end{figure}

This subsection introduces the proposed model to analyze the performance of the OOS system. The developed model leverages queueing theory and inventory control methods; the integrated queueing and inventory model is summarized in Fig. \ref{Integrated}. The inputs to the model include the probability distributions of the travel and repair times ($f_{S_{\text{outbound}}}, f_{S_{\text{repair}}}, f_{S_{\text{inbound}}}$), the individual module failure rate $\alpha$, the number of modules $N$, the length of lead time $L$, the mean launch rate $\beta$, and the required fill rate $\Phi$. The integrated queueing and inventory control model takes those inputs and derives the mean waiting time $\mathbb{E}\left[W\right]$ and the depot capacity $C$. The queueing and the inventory control submodels interact as follows: the queueing submodel generates the spare demand (i.e., module failure) distribution $f_D$, which is fed into the inventory control submodel to calculate the probability distribution of the extra servicing time due to stockout $f_{S_{\text{stockout}}}$; this distribution is fed back into the queueing submodel. Since it is difficult to analytically handle the spare demand distribution $f_D$ due to its state-dependent nature, we approximate this by a Poisson process with the corresponding mean demand rate $\lambda$. In this way, $f_D$ can be characterized using only $\lambda$, which is part of the output from the queueing submodel. (Note that this value is not just the consolidation of the individual module failure rate $\alpha$ because the system failure rate is state-dependent; see Section \ref{queue}.) This approximation is demonstrated to perform well in later numerical simulations. 

The following subsubsections introduce the details of the queueing submodel and inventory control submodel in the integrated model as well as their mathematical coupling.

\subsubsection{Queueing Submodel}
\label{queue}
The queueing part of the OOS problem can be modeled using a finite-source queue. A finite-source queue represents the case where the number of customer satellite modules is finite. Consequently, as the number of failed modules increases, the number of active modules decreases, thus decreasing the system-level failure rate. This fact makes the failure rate state-dependent (i.e., the failure rate depends on the number of active modules), and thus makes the problem challenging.

Using Kendall's notation \cite{Kendall,Lee}, the considered queue is written as $M/G/1/N/N$ (or $M/G/1//N$ depending on the literature \cite{Sztrik2010}). The meaning of each letter is as follows: 
\begin{itemize}
	\item $M$: The arrival process is Markovian (Poisson process).
	\item $G$: The service time distribution is general.
	\item $1$: The number of servicers is one.
	\item $N$: The number of failures allowed in the queue is $N$. 
	\item $N$: The size of the source population is $N$.
\end{itemize}

The general solution for the $M/G/1/N/N$ queue can be found using the Laplace-Stieltjes transform of the service time distribution. Denoting the Laplace-Stieltjes transform of a function $f$ as
$\left\lbrace \mathcal{L}*f\right\rbrace (\theta)$, the mean spare demand rate $\lambda$ becomes:
\begin{equation}
\label{queue1}
\lambda=\frac{1-P_0}{\mathbb{E}\left[ S\right]}
\end{equation}
where
\begin{equation}
\label{queue3}
P_0=\left[ 1+N \mathbb{E}\left[ S\right] \alpha \sum_{n=0}^{N-1}\binom{N-1}{n}B_n\right] ^{-1} 
\end{equation}
\begin{equation}
\label{queue4}
B_n=\begin{cases}
1 &\text{if $n=0$}\\
\prod_{i=1}^{n}\left( \frac{1-\left\lbrace \mathcal{L}*f_S\right\rbrace (i\alpha)}{\left\lbrace \mathcal{L}*f_S\right\rbrace (i\alpha)}\right)  &\text{if $n=1,2,...,N-1$}
\end{cases}
\end{equation}
The mean waiting time $\mathbb{E}\left[W\right]$ is:
\begin{equation}
\label{queue2}
\mathbb{E}\left[W\right]=\frac{N}{\lambda}-\frac{1}{\alpha}-\mathbb{E}\left[S_{\text{inbound}}\right]
\end{equation}
The derivation of the above equations can be found in Refs. \cite{Takacs1962} and \cite{Gupta1994}\footnote{The last term of Eq.~\eqref{queue2} (i.e., the subtraction of $\mathbb{E}\left[S_{\text{inbound}}\right]$) is added to accommodate our definition of the waiting time.}. Here, the mean and the probability distribution of the service time, $\mathbb{E}\left[ S\right]$ and $f_S$, can be found using Eq.~\eqref{serviceTime}, where the distributions of $S_{\text{outbound}}$, $S_{\text{repair}}$, and $S_{\text{inbound}}$ are known and the distribution of $S_{\text{stockout}}$ can be found using the inventory analysis (see Eq.~\eqref{stockout1}). Also, $\lambda$ in Eq.~\eqref{queue1} is used to generate the demand distribution $f_D$ with a Poisson assumption as discussed above:
	\begin{equation}
	f_{D}(i;\lambda,t)=\frac{\lambda^i t^i e^{-\lambda t}}{i!}
	\label{demandDist}
	\end{equation}
This $f_D$ is then fed into the inventory control analysis.

Note that an implicit assumption for this queueing analysis is that, since the servicer's return travel time $S_{\text{inbound}}$ is included as part of the service time, the replaced module does not resume normal operation (i.e., module failure process does not restart) until the servicer returns to the depot after its service. This approximation is reasonable in practice especially when the module's mean time between failures (MTBF) is sufficiently long, as demonstrated in the later comparison with simulations.

\subsubsection{Inventory Control Submodel}
The inventory control analysis for the OOS problem contains two parts: finding $C$ given $\Phi_{\text{req}}$ and finding $f_{S_{\text{stockout}}}$.

For the order-up-to policy \cite{Cachon}, the fill rate $\Phi$ is defined as follows:
\begin{equation}
\label{inventoryPhi}
\begin{split}
\Phi &=1-\frac{\int_0^\infty \sum_{i=1}^\infty \max (0,i-C) f_D (i;\lambda,t+L)f_{T_\text{l}}(t;\beta) dt}{\int_0^\infty \sum_{i=1}^\infty i f_D (i;\lambda,t)f_{T_\text{l}}(t;\beta) dt}\\
& =1-\frac{\beta}{\lambda} \int_0^\infty \sum_{i=1}^\infty \max (0,i-C) f_D (i;\lambda,t+L)f_{T_\text{l}}(t;\beta) dt
\end{split}
\end{equation}
where the exponential launch interval distribution is as follows:
	\begin{equation}
	\label{fTl}
	f_{T_\text{l}}(t;\beta)=\beta e^{-\beta t}
	\end{equation}
and the demand distribution $f_D$ can be found in Eq.~\eqref{demandDist}. The numerator of the second term in Eq.~\eqref{inventoryPhi} is \textit{the expected backorder over a launch interval and a lead time}, whereas the denominator is \textit{the expected demand over a launch interval}. The reason why the numerator considers both the launch interval and the lead time (instead of only the launch interval) can be intuitively explained with the following example. Consider a case where an order is made at a point when the inventory level is $I$ units, and there is no other replenishment order on the way. In this case, the order would be delivered after $L$ time steps when the inventory level is $I-I_\text{L}$ units, where $I_\text{L}$ is the additional units of demand during the lead time $L$. Because the order only delivers $C-I$ units using the information when the order was made, the actual inventory right after the replenishment delivery is $(I-I_\text{L})+(C-I)=C-I_\text{L}$ instead of the full capacity $C$. Therefore, a stockout (and thus a backorder) happens when the demand between this delivery and the next delivery (i.e., over one launch interval), denoted as $I_\text{I}$, exceeds $C-I_\text{L}$; this corresponds to when the summation of the demand over an interval ($I_\text{I}$) and the demand over the lead time ($I_\text{L}$) exceeds $C$. This explains why the fill rate computation needs to take into consideration the backorder over both the launch interval and the lead time. For a more rigorous derivation of the fill rate expression in Eq.~\eqref{inventoryPhi}, see Ref. \cite{Cachon}.

A closed-form expression for $\Phi$ (Eq.~\eqref{inventoryPhi}) can be derived as follows:
\begin{align*}
\Phi&=1-\frac{\beta}{\lambda} \int_0^\infty \sum_{i=1}^\infty \max (0,i-C) f_D (i;\lambda,t+L)f_{T_\text{l}}(t;\beta) dt\\
&=1-\frac{\beta}{\lambda} \int_0^\infty \sum_{i=1}^\infty i \beta e^{-\beta t} \frac{\lambda^{C+i}(t+L)^{C+i}e^{-\lambda(t+L)}}{(C+i)!} \, dt \\
&=1-\frac{\beta}{\lambda} \left[ e^{- \lambda L}\left( \frac{\lambda}{\beta+\lambda}\right)^{C} \frac{\lambda }{\beta} \sum_{j=0}^C \frac{  \left\lbrace (\beta+\lambda)L\right\rbrace ^j}{j!} + L\lambda \left\lbrace 1 - e^{- \lambda L}\sum_{j=0}^{C-1}  \frac{(\lambda L)^j}{j!}\right\rbrace  + \left( \frac{\lambda}{\beta} - C\right)\left\lbrace  1 -  e^{-\lambda L}\sum_{j=0}^C  \frac{(\lambda L)^j}{j!}\right\rbrace \right]  
\end{align*}
This expression allows us to evaluate the fill rate $\Phi$ given a depot capacity $C$.
With this expression, we can find the minimum $C$ that satisfies a given fill rate requirement $\Phi_{\text{req}}$:
\begin{align}
\min  \quad & C 
\label{PhiReq1}\\
\text{s.t.} \quad &\Phi \geq \Phi_{\text{req}}
\label{PhiReq2}
\end{align}
Since $\Phi$ is a monotonically increasing function of $C$, the solution to this problem can be found simply by iteratively incrementing $C$ until $\Phi \geq \Phi_{\text{req}} $ is satisfied. 

Using the depot capacity $C$, we can derive the expression for the additional service time due to stockout $S_{\text{stockout}}$, which is then fed back to the queueing analysis. $S_{\text{stockout}}$ corresponds to the extra "service" time that the first repair finding the depot to be out of stock needs to wait before the servicer can depart for the repair service. The impact of this extra service time then propagates to the remaining repairs through their waiting times in the queue $W_q$. To derive the expression for $S_{\text{stockout}}$, we consider the following two cases. If the time it takes to have $C+1$ units of demand (i.e., failures), denoted as $T_\text{s}$, is the same or longer than the sum of the launch interval and the lead time ($T_\text{l}+L$), then no extra service time is added to the repairs in that interval. If that is not the case ($T_\text{l}+L>T_\text{s}$), then we add an extra service time $S_{\text{stockout}}$ corresponding to $T_\text{l}+L-T_\text{s}$ to the first unit of repair demand that finds the depot to be out of stock. Considering that only one out of $\lambda \frac{1}{\beta}$ units of repair demand on average in each launch interval can potentially be affected, $S_{\text{stockout}}$ can be written as follows:  
\begin{equation}
\label{stockout1}
S_{\text{stockout}} =\begin{cases}
\max (T_\text{l}+L-T_\text{s},0) &\text{with $p=\frac{\beta}{\lambda}$}\\
0 &\text{with $p=1-\frac{\beta}{\lambda}$}
\end{cases}
\end{equation}
where $T_\text{l}$ and $T_\text{s}$ themselves are also random variables following the probability density functions $f_{T_\text{l}}(t;\beta)$ and $f_{T_\text{s}}(t;\lambda,C+1)$, respectively. $f_{T_\text{l}}(t;\beta)$ can be found using Eq.~\eqref{fTl}, and $f_{T_\text{s}}(t;\lambda,C+1)$ can be expressed with an Erlang distribution because of the Poisson demand approximation:
\begin{equation}
f_{T_\text{s}}(t;\lambda,C+1)=\frac{\lambda^{C+1} t^{C} e^{-\lambda t}}{C!}=f_D(C;\lambda,t)\lambda
\end{equation}
Note that this is a conservative approximation; in reality, $S_{\text{stockout}}$ can be shorter because the impact of stockout does not take effect until all the ongoing repairs in the queue with the in-stock spares are completed.

One comment about the assumption behind the inventory control submodel needs to be added: the considered model does not have an upper limit on the number of units a rocket can carry even when there are a large number of backorders; although we have a rocket capacity constraint in reality, this assumption is reasonable when the solutions are near the knee region, where a stockout happens rarely. In later simulations, a rocket capacity constraint up to $C$ units is enforced, and the proposed model is demonstrated to approximate the simulation results well nevertheless.

\subsubsection{Coupling between the Queueing and Inventory Control Submodels}
We next look at how the queueing submodel and the inventory control submodel are mathematically coupled. 

First, we consider the queueing submodel and derive the expressions for $\lambda$ and $\mathbb{E}\left[W\right]$ for a given distribution of $S_{\text{stockout}}$. Examining Eqs.~\eqref{queue1}-\eqref{queue2}, we can see that the only complication for this process is obtaining the expressions for $\mathbb{E}\left[S\right]$ and $\left\lbrace \mathcal{L}*f_S\right\rbrace (\theta)$. $\mathbb{E}\left[S\right]$ can be expressed as:
\begin{equation}
\label{sum}
\mathbb{E}\left[S\right] = \mathbb{E}\left[S_{\text{stockout}}\right]+\mathbb{E}\left[S_{\text{outbound}}\right]+\mathbb{E}\left[S_{\text{repair}}\right]+\mathbb{E}\left[S_{\text{inbound}}\right]
\end{equation}
Similarly, leveraging the convolution theorem for the Laplace-Stieltjes transform, $\left\lbrace \mathcal{L}*f_S\right\rbrace (\theta)$ can be expressed as:
\begin{equation}
\label{LS}
\left\lbrace \mathcal{L}*f_S\right\rbrace (\theta) = \left\lbrace \mathcal{L}*f_{S_{\text{stockout}}}\right\rbrace (\theta)\left\lbrace \mathcal{L}*f_{S_{\text{outbound}}}\right\rbrace (\theta) \left\lbrace \mathcal{L}*f_{S_{\text{repair}}}\right\rbrace (\theta) \left\lbrace \mathcal{L}*f_{S_{\text{inbound}}}\right\rbrace (\theta) 
\end{equation}
The terms for $S_{\text{outbound}}$, $S_{\text{repair}}$, and $S_{\text{inbound}}$ in Eqs.~\eqref{sum}-\eqref{LS} can be derived using orbital mechanics; deriving this analytical expression is trivial because we know the exact locations of the customer satellites. Thus, if we are given the closed-form expressions for $\mathbb{E}\left[S_{\text{stockout}}\right]$ and $\left\lbrace \mathcal{L}*f_{S_{\text{stockout}}}\right\rbrace (\theta)$ from the inventory control submodel, $\lambda$ and $\mathbb{E}\left[W\right]$ can be found analytically.

Next, we examine the terms for $S_{\text{stockout}}$ in Eqs.~\eqref{sum}-\eqref{LS}, $\mathbb{E}\left[S_{\text{stockout}}\right]$ and $\left\lbrace \mathcal{L}*f_{S_{\text{stockout}}}\right\rbrace (\theta)$, through the inventory control submodel. These terms can be analytically expressed using $C$ (obtained from Eqs.~\eqref{PhiReq1}-\eqref{PhiReq2}) and $\lambda$. We first derive $\left\lbrace \mathcal{L}*f_{S_{\text{stockout}}}\right\rbrace (\theta)$ and use that to find $\mathbb{E}[S_{\text{stockout}}]$. 
\begin{equation}
\left\lbrace \mathcal{L}*f_{S_{\text{stockout}}}\right\rbrace (\theta)
=\frac{\beta}{\lambda}\left\lbrace \mathcal{L}*f_{\max (T_\text{l}+L-T_\text{s},0)}\right\rbrace (\theta)+\left(1-\frac{\beta}{\lambda}\right)
\end{equation}
Focusing on $\left\lbrace \mathcal{L}*f_{\max (T_\text{l}+L-T_\text{s},0)}\right\rbrace (\theta)$,
\begin{align}
&\left\lbrace \mathcal{L}*f_{\max (T_\text{l}+L-T_\text{s},0)}\right\rbrace (\theta)\nonumber \\
&= \int_0^\infty \left( \int_0^{t_\text{l}+L} e^{-\theta (t_\text{l}+L-t_\text{s})} \beta e^{-\beta t_\text{l}} \frac{\lambda^{C+1} t_\text{s}^C e^{-\lambda t_\text{s}}}{C!} \, dt_\text{s} \right) dt_\text{l} + \int_0^\infty \left( \int_{t_\text{l}+L}^{\infty} \beta e^{-\beta t_\text{l}} \frac{\lambda^{C+1} t_\text{s}^C e^{-\lambda t_\text{s}}}{C!} \,  dt_\text{s} \right) dt_\text{l} \nonumber  \\
&=\frac{\beta}{\beta+\theta}\left(\frac{\lambda}{\lambda-\theta}\right)^{C+1} e^{-\theta L} \left[ 1-\sum_{n=0}^C e^{-(\lambda-\theta)L} \frac{(\beta+\theta)(\lambda-\theta)^n}{(\beta+\lambda)^{n+1}}\sum_{i=0}^n\frac{\left\lbrace (\beta+\lambda)L\right\rbrace ^i}{i!}\right] +\sum_{n=0}^C e^{-\lambda L}\frac{\beta \lambda^n}{(\beta+\lambda)^{n+1}}  \sum_{i=0}^n\frac{\left\lbrace (\beta+\lambda) L\right\rbrace ^i}{i!}
\label{stockoutLaplace}
\end{align}
Thus, we have obtained a closed-form expression for $\left\lbrace \mathcal{L}*f_{S_{\text{stockout}}}\right\rbrace (\theta)$.
Using $\left\lbrace \mathcal{L}*f_{S_{\text{stockout}}}\right\rbrace (\theta)$, $\mathbb{E}[S_{\text{stockout}}]$ can be found as follows:
\begin{align}
&\mathbb{E}[S_{\text{stockout}}]\nonumber \\
&=-\frac{d}{d\theta}\left\lbrace \mathcal{L}*f_{S_{\text{stockout}}}\right\rbrace (\theta)|_{\theta=0}\nonumber \\
&=\frac{\beta}{\lambda}\left[ L+\frac{1}{\beta}-\frac{C+1}{\lambda}+\sum_{n=0}^C (C+1-n) e^{-\lambda L} \frac{\beta \lambda^{n-1}}{(\beta+\lambda)^{n+1}}\sum_{i=0}^n \frac{\left\lbrace (\beta+\lambda)L\right\rbrace ^i}{i!} \right] 
\label{stockoutExp}
\end{align}

As we can see in the resulting expressions, the terms needed for the expressions of $\lambda$ and $\mathbb{E}\left[W\right]$ depend on the distribution of $S_\text{stockout}$ (i.e., through the queueing analysis), and this distribution of $S_\text{stockout}$ depends on $\lambda$ (i.e., through the inventory control analysis). Therefore, this system of equations needs to be solved concurrently\footnote{A comment on the practical implementation: some terms in the derived equations can span many orders of magnitude (e.g., the first term of Eq.~\eqref{stockoutLaplace}), which can potentially cause numerical instability. One technique to avoid such an issue is to implement the multiplications in those terms as additions in the log-domain.}.

\subsection{Solution Method} 
With the proposed coupled queueing and inventory control submodels, our goal is to find the solution of the mean waiting time $\mathbb{E}\left[W\right]$ and the depot capacity $C$ for a given fill rate requirement $\Phi_{\text{req}}$ that is close to one (i.e., around the knee region of the $\mathbb{E}\left[W\right]$ vs. $C$ curve). The coupled equations for these two submodels generally cannot be solved fully analytically; a standard numerical solver such as the \verb|fsolve| function in MATLAB can be leveraged. The solution of these coupled equations enables us to find both $\mathbb{E}\left[W\right]$ (i.e., from the queueing analysis) and $C$ (i.e., from the inventory analysis) for a given $\Phi_\text{req}$.

The performance of solving a set of coupled equations depends on the initial value. As the initial value for solving our problem, we propose to use the no-stockout case, i.e., by setting $S_{\text{stockout}}=0$. In this case, $f_{S_\text{stockout}}$ is not needed for the queueing analysis, and therefore the loop between the queueing and inventory control submodels is decoupled. Thus, we can independently find the mean waiting time $\mathbb{E}\left[W\right]$ and mean demand rate $\lambda$ via Eqs.~\eqref{queue1}-\eqref{queue2}. For highly reliable realistic OOS applications (i.e., $\Phi_{\text{req}}\approx 1$), this solution is expected to be close from the true solution, and thus serves as a good initial value for the \verb|fsolve| function.

\section{Application Example}
\label{simulations}
This section applies the proposed method to an example case with realistic parameters and assesses the accuracy of the proposed model with simulations. The considered example case contains 10 customer satellites with 5 modules each that are evenly distributed over the geosynchronous orbit, and the depot is collocated with one of these satellites. The outbound and inbound travel times are computed based on the phasing maneuver between the depot and the satellite with the failed module, and the operations needed between each repair trip (e.g., loading a new spare, refueling) are assumed to be near-instantaneous\footnote{The spare loading and refueling operations are assumed as near-instantaneous while the repair operations are not because the former are generally routine operations with a cooperative target (i.e., depot), whereas the latter include additional dedicated and involved operations with a non-cooperative target (i.e., failed customer satellites). This assumption can be easily relaxed by including additional corresponding service times.}. See Appendix for the specific phasing maneuver strategy chosen in this example. The values of the key parameters are listed in Table \ref{tab:table1}. Three cases for the module MTBF are considered.  
\begin{table}[hbt!]
\caption{\label{tab:table1} Simulation Parameters. MTBF stands for the mean time between failures.}
\centering
\begin{tabular}{lcr}
\hline
\hline
\multicolumn{2}{c}{Parameter}   & \multicolumn{1}{c}{Value}\\\hline
Repair time & $S_{\text{repair}}$ & 4 hours\\
Launch lead time &$L$ & 2160 hours \\
Mean launch interval & $1/\beta$ & 1213.4 hours \\
MTBF per module & $1/\alpha$ & {{20000, 10000, 4000}} hours\\
\hline
\hline
\end{tabular}
\end{table}

For the considered application case, the developed semi-analytical model is used to derive the mean waiting time vs. depot capacity curve\footnote{For this illustrative example, the \verb|fsolve| function in MATLAB is used with the function tolerance of $10^{-4}$, the optimality tolerance of $10^{-4}$, and the step tolerance of $10^{-4}$.}. To assess the accuracy of the proposed semi-analytical model, we use the depot capacity derived from the analysis and evaluate the corresponding waiting time of the system via simulations; the mean waiting time results from the simulations are compared with those from the semi-analytical model. For each given set of the MTBF and the depot capacity, 500 simulation runs are performed over a time horizon of 200000 hours (i.e., $\sim 22.8$ years). The simulation methods used in this paper are based on Ref. \cite{SartonduJonchay2017}. As discussed previously, to reflect the reality, the simulations enforce a rocket capacity constraint that prevents the replenishment rocket from delivering more units than the depot capacity even when there are backorders. We evaluate the cases with fill rate requirements $\Phi_{\text{req}}=0.8, 0.85, 0.9, 0.95, 0.99, 0.995, 0.999$, and expect the proposed model to perform well (i.e., approximate the simulation results accurately) when $\Phi_{\text{req}}$ is close to one. 

\begin{figure}[h!]
\centering
\includegraphics[trim=0 0 40 10,clip,scale=0.75]{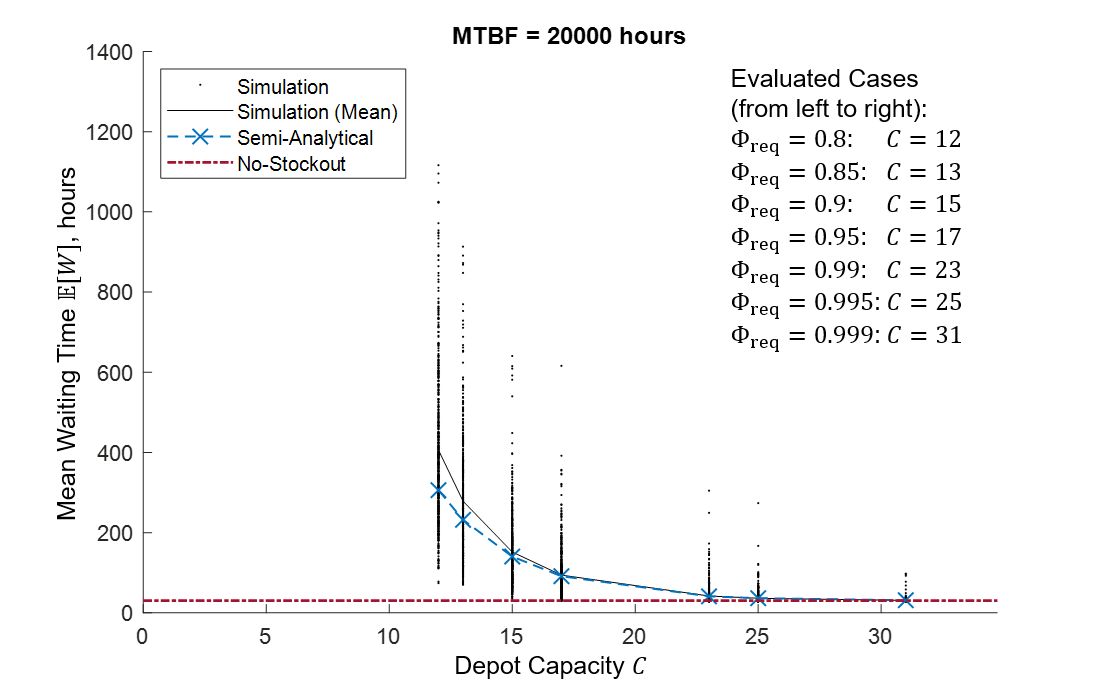}
\caption{Mean waiting time vs. depot capacity for mean time between failure = 20000 hours. The cases with various fill rate requirements $\Phi_{\text{req}}=0.8, 0.85, 0.9, 0.95, 0.99, 0.995, 0.999$ are shown (from left to right) as crosses. The no-stockout case is also shown for reference.}
\label{Formulation 12a}
\end{figure}

\begin{figure}[h!]
\centering
\includegraphics[trim=0 0 40 10,clip,scale=0.75]{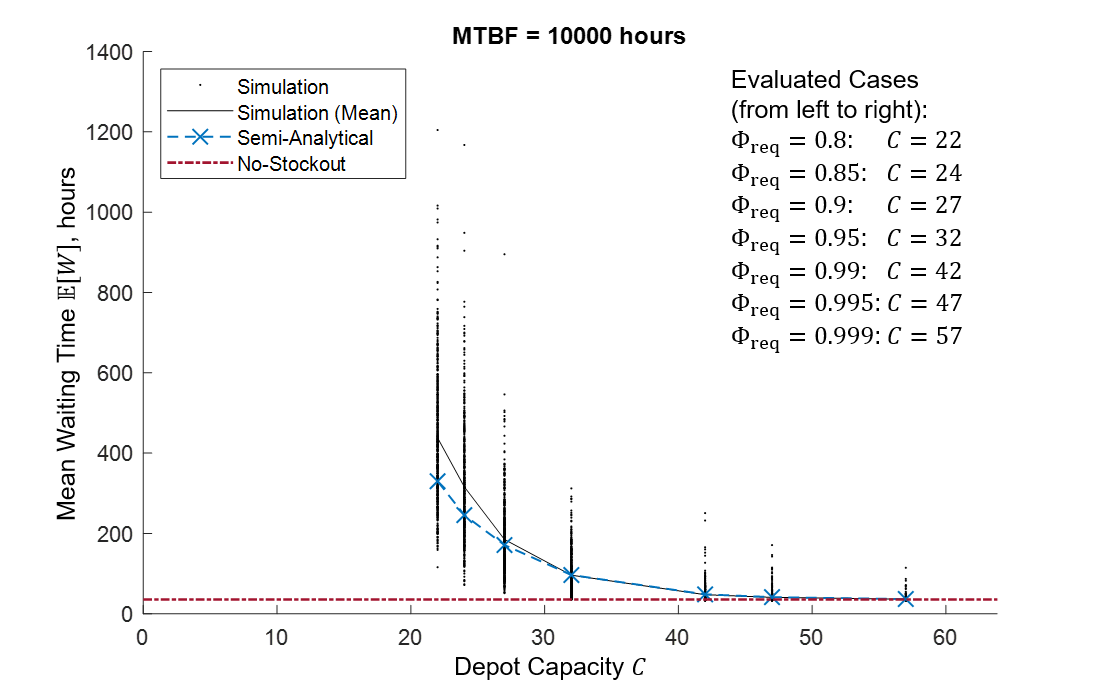}
\caption{Mean waiting time vs. depot capacity for mean time between failure = 10000 hours. The cases with various fill rate requirements $\Phi_{\text{req}}=0.8, 0.85, 0.9, 0.95, 0.99, 0.995, 0.999$ are shown (from left to right) as crosses. The no-stockout case is also shown for reference.}
\label{Formulation 12b}
\end{figure}

\begin{figure}[h!]
\centering
\includegraphics[trim=0 0 40 10,clip,scale=0.75]{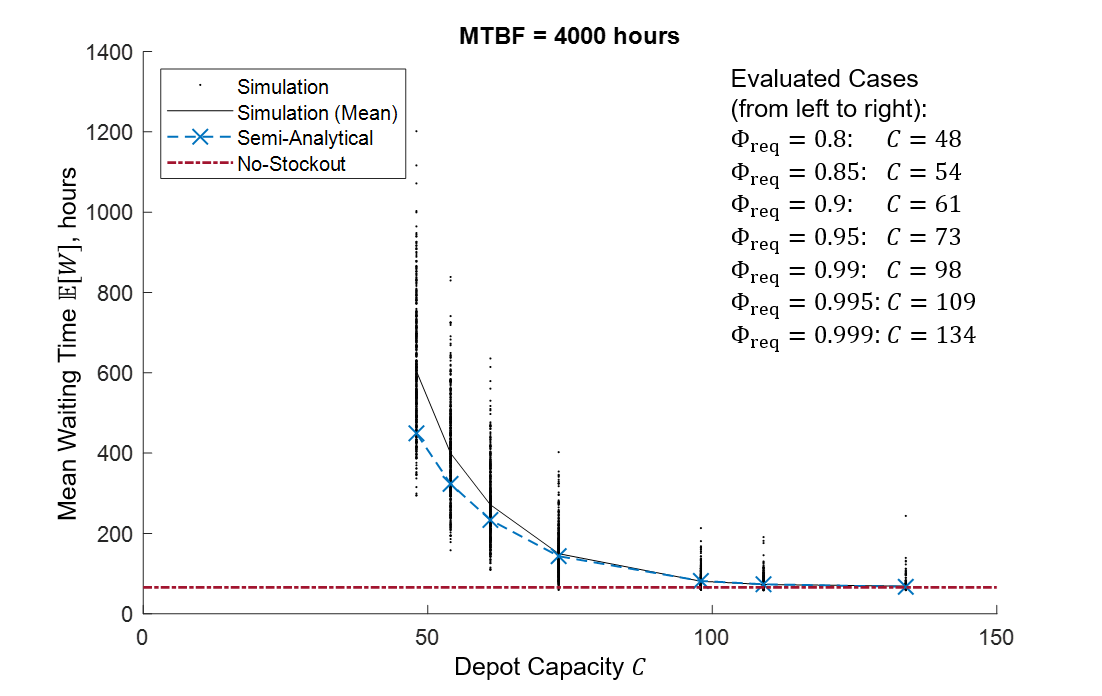}
\caption{Mean waiting time vs. depot capacity for mean time between failure = 4000 hours. The cases with various fill rate requirements $\Phi_{\text{req}}=0.8, 0.85, 0.9, 0.95, 0.99, 0.995, 0.999$ are shown (from left to right) as crosses. The no-stockout case is also shown for reference.}
\label{Formulation 12c}
\end{figure}

The results of the semi-analytical model and the simulations are shown in Figs. \ref{Formulation 12a}-\ref{Formulation 12c}. For the simulation results, the averaged waiting times over the time horizon for each 200000-hour run are evaluated for the corresponding depot capacities and are shown as individual black dots in Figs. \ref{Formulation 12a}-\ref{Formulation 12c}; additionally, their means over the 500 simulations for each depot capacity are connected by a black solid line. The semi-analytical results for each fill rate are shown as crosses connected by a blue dashed line in Figs. \ref{Formulation 12a}-\ref{Formulation 12c}. In addition, we derive the mean waiting time with no stockout (i.e., infinite capacity), shown as a horizontal red dash-dot line in Figs. \ref{Formulation 12a}-\ref{Formulation 12c}.

The quantitative comparison of the results from the model and simulations is also shown in Table \ref{tab:TableKnee2}. Note that the error between the model and the simulations is caused by both the approximation made in the model and the randomness in the simulations due to the finite number of runs.

\begin{table}[h!]
\caption{\label{tab:TableKnee2} Comparison between the semi-analytical results $W_\text{semi-analytical}$ and the simulation results $W_\text{simulation}$. The no-stockout mean waiting time is also included for reference. The error in $\mathbb{E}\left[W\right]$ is evaluated as $\left|\left(\mathbb{E}\left[W_\text{semi-analytical}\right]-\mathbb{E}\left[W_\text{simulation}\right]\right)/\mathbb{E}\left[W_\text{simulation}\right]\right|$.}
\centering
\begin{tabular}{rrrrrr}
\hline
\hline
MTBF, hours &  $\Phi_\text{req}$  & $C$ &  $\mathbb{E}\left[W_\text{semi-analytical}\right]$ hours & $\mathbb{E}\left[W_\text{simulation}\right]$, hours & Error in $\mathbb{E}\left[W\right]$ \\\hline
\multirow{8}{*}{20000} & 0.8 & 12 & 306.1 & 406.4 & 24.7\%\\
& 0.85 & 13 & 232.1 & 278.5 & 16.7\%\\
& 0.9 & 15 & 140.5 & 151.9 & 7.5\%\\
& 0.95 & 17 & 91.5 & 94.2 & 2.9\%\\
& 0.99 & 23 & 41.3 & 42.1 & 2.0\%\\
& 0.995 & 25 & 36.6 & 36.6 & 0.0\%\\
& 0.999 & 31 & 31.6 & 31.4 & 0.8\%\\
& no-stockout & -- & 30.5 & -- & --\\\hline
\multirow{8}{*}{10000} & 0.8 & 22 & 330.1 & 437.9 & 24.6\%\\
& 0.85 & 24 & 245.7 & 315.9 & 22.2\%\\
& 0.9 & 27 & 171.3 & 185.0 & 7.4\%\\
& 0.95 & 32 & 96.6 & 96.0 & 0.5\%\\
& 0.99 & 42 & 48.4 & 47.5 & 1.8\%\\
& 0.995 & 47 & 41.4 & 40.9 & 1.4\%\\
& 0.999 & 57 & 36.8 & 36.5 & 0.6\%\\
& no-stockout & -- & 35.5 & -- & --\\\hline
\multirow{8}{*}{4000} & 0.8 & 48 & 449.4 & 603.9 & 25.6\%\\
& 0.85 & 54 & 323.5 & 399.9 & 19.1\%\\
& 0.9 & 61 & 233.6 & 271.6 & 14.0\%\\
& 0.95 & 73 & 143.4 & 149.7 & 4.2\%\\
& 0.99 & 98 & 81.7 & 81.0 & 0.9\%\\
& 0.995 & 109 & 73.5 & 72.9 & 0.8\%\\
& 0.999 & 134 & 67.2 & 68.8 & 2.3\%\\
& no-stockout & -- & 65.8 & -- & --\\
\hline
\hline
\end{tabular}
\end{table}

From the results, we can observe that the proposed method explores the relationship between the mean waiting time and the depot capacity around the knee region. Its evaluation of the mean waiting time achieves an accuracy of $<5\%$ when $\Phi_{\text{req}}\geq 0.95$. 
Note that, in practice, the OOS system would be designed to have a high fill rate (i.e. $\Phi_{\text{req}}\geq 0.95$) in order to achieve reasonable waiting times for service; therefore, the approximation in the developed model serves a purpose. Additionally, we can also observe that the no-stockout solution serves as an optimistic lower bound on the mean waiting time. This no-stockout case corresponds to an ideal OOS system with infinite depot capacity and can be used as a first-order approximation of the mean waiting time. Overall, the results demonstrate the accuracy and utility of the proposed semi-analytical model for the considered OOS application. 

One benefit of the proposed model is that it does not require computationally costly simulations for evaluation. For the considered example, the simulations take more than 15 hours to complete the evaluation of all cases with Python 3.6, whereas the semi-analytical model only takes approximately 5 seconds in total with MATLAB R2019a\footnote{All tests were performed on an Intel Core i7-8650U CPU @ 1.90GHz platform with 16GB RAM.}. Although the exact computational time depends on the implementation details, the substantial computational cost saving provided by the developed model is evident. The developed model provides an efficient high-level design analysis and optimization method with sufficient accuracy for early-stage systems design, complementing the existing costly simulation techniques for detailed design.

\section{Conclusion}
\label{conclusion}
This paper develops a novel semi-analytical model for OOS system analysis based on queueing theory and inventory management methods. We consider an OOS system that provides responsive services to the random failures of a number of modular customer satellites in orbit. The considered OOS architecture comprises a servicer that provides module-replacement services to the customer satellites, an on-orbit depot that stores the spare modules, and a series of launch vehicles to refill the depot. The developed model is capable of analyzing the queue of the service operations as well as the logistics of the spares, evaluating the mean waiting time before service completion for a given failure and its relationship with the depot capacity. The case study shows that the results from the semi-analytical model approximate the solution well without computationally costly simulations.

Although this paper uses a simple OOS application case to demonstrate the value of the developed model, the developed general approach can be applied to more complex OOS problems with reasonable modifications. Possible extensions include: (1) employing a servicer that can carry multiple spares (i.e., servicing multiple customers in one trip); (2) considering multiple types of spares, repair operations, or even servicer spacecraft; and (3) applying alternative inventory control policies for the depot. We expect that this paper opens up a broad range of applications of semi-analytical models to OOS system design.

\section*{Acknowledgment}
The authors thank Brian Hardy for running the simulations needed for this paper and Hang Woon Lee and Onalli Gunasekara for reviewing this paper.

\section*{Funding Sources}
This work is partially supported by the Defense Advanced Research Projects Agency (DARPA) Young Faculty Award D19AP00127. The content of this paper does not necessarily reflect the position or the policy of the U.S. Government, and no official endorsement should be inferred. Approved for public release; distribution is unlimited.

\section*{Appendix: Assumption on Phasing Maneuver in the Application Example }
This appendix summarizes the assumption on the phasing maneuvers at the geosynchronous orbit used in the example in this paper. Note that although this particular maneuver strategy is chosen as an example, the proposed method is compatible with other maneuver strategies as well.

The travel time for each phasing maneuver over an angular separation $\Delta \Theta$ (defined as the initial phasing angle of the servicer with respect to the target) at a circular orbit of radius $r_\text{target}$ ($\sim 42160$ km in the case of the geosynchronous orbit) is determined in the following way. We first define two integer parameters $k_1\geq 1$ and $k_2\geq 0$, where $k_1$ is the number of orbits the servicer travels in the phasing orbit and $k_2$ is the number of orbits the target travels in its circular orbit before the rendezvous. Using the definition of $k_2$, the travel time can be expressed as:
\begin{equation} \label{phasing time}
\ {t_\text{travel}} = (\Delta \Theta + 2\pi k_2) \sqrt{\frac{r_\text{target}^3}{GM_\text{E}}}
\end{equation}
where $G$ is the universal gravitational constant and $M_\text{E}$ is the mass of the Earth. Furthermore, using the definition of $k_1$, we can find the relationship between the travel time and the semimajor axis of the phasing orbit $a$:
\begin{equation} \label{phasing time2}
\ {t_\text{travel}} = 2\pi k_1 \sqrt{\frac{a^3}{GM_\text{E}}}
\end{equation}
Combining Eqs.~\eqref{phasing time}-\eqref{phasing time2} to an expression for $a$ using $k_1$ and $k_2$:
\begin{equation}
a=\left(\frac{\Delta \Theta + 2\pi k_2}{2\pi k_1}\right)^\frac{2}{3}r_\text{target} 
\end{equation}

In the considered example, the values of $k_1$ and $k_2$ are chosen so that they minimize the travel time while satisfying the following requirement:
\begin{equation}\label{critical}
a\geq \frac{r_\text{target}+R_\text{E}+h_\text{crit}}{2}
\end{equation}
where $R_\text{E}$ is the radius of the Earth, and $h_\text{crit}$ is the minimum altitude for the phasing orbit. In the numerical example, $h_\text{crit}=10000$ km is used. This minimum altitude is applied to ensure minimum interference of the servicer with the environment at lower orbits (e.g., atmospheric drag, orbital debris, other satellites). Note that the minimum travel time indicates the minimum feasible $k_2$ according to Eq.~\eqref{phasing time}; therefore, our goal is equivalent to finding the minimum possible integer $k_2 \geq 0$ that yields a feasible integer $k_1\geq 1$ with respect to the constraint in Eq.~\eqref{critical}. Such a solution can be found by using a simple iterative process.

\bibliography{References}

\end{document}